\documentclass{aa}
\usepackage{graphicx}

\begin{document}

\title{Multiband optical polarimetry of BL Lac objects with the Nordic Optical
Telescope\thanks{Based on observations made with the Nordic Optical Telescope,
operated on the island of La Palma jointly by Denmark, Finland,
Iceland, Norway, and Sweden, in the Spanish Observatorio del
Roque de los Muchachos of the Instituto de Astrofisica de
Canarias. } }
\author{L. Tommasi \inst{1}
\and E. Palazzi \inst{2}
\and E. Pian \inst{3}
\and V. Piirola \inst{4}
\and E. Poretti \inst{5}
\and F. Scaltriti \inst{6}
\and A. Sillanp\"a\"a \inst{4}
\and L. Takalo \inst{4}
\and A. Treves \inst{7}}

\offprints{Elena Pian, \email{pian@ts.astro.it}}

\institute{Dipartimento di Fisica, Universit\`a di Milano,
I-20133 Milano, Italy \and ITESRE-CNR, I-40129 Bologna, Italy
\and Osservatorio Astronomico di Trieste, I-34131 Trieste, Italy
(also ITESRE-CNR, I-40129 Bologna, Italy) \and Tuorla
Observatory, SF-21500 Piikkio, Finland \and Osservatorio
Astronomico di Brera, I-23807 Merate, Italy \and Osservatorio
Astronomico di Torino, I-10025 Pino Torinese, Italy \and
Dipartimento di Scienze, Universit\`a dell'Insubria, I-22100
Como, Italy}

\date{Received 3 January 2001 / Accepted 8 June 2001}

\markboth{L. Tommasi et al.:  Multiband optical polarimetry of BL Lac objects with the
Nordic Optical Telescope}{}

\abstract{ Optical polarization of seven selected BL Lac objects
in \mbox{$UBVRI$} bands was studied with the Nordic Optical
Telescope from December 10-14, 1999. Two of them, \object{3C 66A}
and \object{PKS~0735+178}, were monitored for 4 nights for a
total integration time of 4.75 and 5.5 hours, respectively. Other
objects (\object{1Jy~0138$-$097}, \object{H~0414+009},
\object{PKS~0823$-$223}, \object{OJ287} and \object{BL Lac}) were
observed sparsely during the run. Apart from PKS~0823$-$223 (more
polarized than observed in the past), the sources show levels of
flux and polarization consistent with results at previous epochs.
3C 66A and PKS~0735+178 were intensively observed during December
11 and 12 and exhibited variability of polarization, both on
internight and intranight time scales. Wavelength dependence of
polarization has been investigated, as well as circular
polarization. The results are discussed within the standard model
for BL Lacs. }

\maketitle

\section{Introduction}

Synchrotron  radiation from a relativistic jet pointing toward
the observer is thought to produce the emission of BL Lac
objects, from radio up to X-rays and sometimes soft gamma rays.
The maximum power output of the synchrotron component is located
at different energies in different objects. In particular,
following Padovani and Giommi (\cite{Padovani95}), BL Lacs can be
divided into two classes: Low energy cutoff BL Lacs (LBL) and
High energy cutoff BL Lacs (HBL). In the former, the synchrotron
peak, in a $\nu F_{\nu}$ representation, lies in the
optical/infrared domain, while in the latter it occurs in the
UV/X-ray band. A second, higher energy peak, occurring in the
GeV/TeV domain for LBL/HBL, respectively, is due to inverse
Compton radiation. The strongest observational evidence in favour
of synchrotron radiation is the high polarization recorded in all
the bands from radio to UV (Brindle et al. \cite{Brindle}; Saikia
\& Salter \cite{Saikia88}; Allen et al. \cite{Allen}; Gabuzda et
al. \cite{Gabuzda96}). This is related to the presence of a
regular (at least in one of its components) magnetic field. The
relativistic beaming amplifies the radiation from the nucleus
with respect to the unpolarized light coming from the host galaxy
and the polarization can be as high as 30\%. Only a few objects
have been observed up to now in search for polarization
variability on intranight timescales (Puschell et al.
\cite{Puschell79}; Moore et al. \cite{Moore87}; Impey et al.
\cite{Impey00}; Tommasi et al. \cite{Tommasi01}).

Data presented in this paper have been collected with the Nordic
Optical Telescope (NOT) during December 10-14, 1999, within a
programme started some years ago and devoted to the optical
polarization monitoring of BL Lac objects in both
hemispheres, in a range of time scales from days down to less than an hour
(Scaltriti et al. \cite{Scaltriti99}; Treves et al.
\cite{Treves99}; Tommasi et al. \cite{Tommasi01}).

A description of the observations and data reduction is given in
Sect.~2; the results of linear and circular polarimetry are
reported in Sect.~3 and discussed in Sect.~4. Preliminary and
partial results were presented in Tommasi et al.
(\cite{TommasiTS}).

\section{Observations}
We selected 11 northern BL Lacs among the brightest and most
polarized objects observed by Impey \& Tapia (\cite{Impey88}) and
Takalo et al. (\cite{Takalo92}). During the first night of the
run, we performed a brief photopolarimetric ``survey" with short
integration times on all the sources, to determine which of them
were the brightest and most polarized at the epoch of
observation, to be intensively monitored in the two following
nights: 3C 66A and PKS~0735+178 were chosen. The total effective
integration times on these two objects were 4.75 and 5.5 hours,
respectively. Five more BL Lacs included in the list were
observed sparsely during the run: 1Jy~0138$-$097 (only circular
polarization), H~0414+009, PKS~0823$-$223, OJ~287, BL Lac. In
Table~\ref{sample}
the list of the observed sources is reported along with some
results and basic information (see following sections), together
with three southern BL Lacs monitored by our team with the 2.15~m
telescope of the Complejo Astronomico El Leoncito (CASLEO,
Argentina; see Scaltriti et al. \cite{Scaltriti99};
Tommasi et al. \cite{TommasiTS}; Treves et
al. \cite{Treves99}).
\begin{table*}
\label{sample}
 \centering
\begin{minipage}{400pt} \caption[]{Summary of the relevant parameters of
the observed BL
Lacs.} \centering \footnotesize
\begin{tabular}{lccrcrrccc} \hline            
\rule[-0pt]{0pt}{11pt} 
{\em Object}& {\em Catalog, name} & {\em Type} &
\multicolumn{1}{c}{\em z} & {\em m$_V$} &
\multicolumn{1}{c}{$P_m$ \footnote{Average value of linear
polarization in the $V$ band. For the objects observed in
different epochs, the average absolute value of polarization
percentage is given, neglecting the position angle.}} &
\multicolumn{1}{c}{$P_\nu$ \footnote{Wavelength dependence of
polarization, as defined in Eq.~\ref{pnu}. For objects observed
in more than one night, the weighted average is given. }} & {\em
CD} \footnote{Ratio between nuclear and galactic flux in $R$ band
(Wurtz et al. 1996, Scarpa et al. 2000).}& $B$ \footnote{Magnetic
field intensity in Gauss (Ghisellini et al. 1998, Tagliaferri et
al. 2001).}& $\delta$ \footnote{Doppler factor (Ghisellini
et al. 1998, Tagliaferri et al. 2001).}\\
\hline
0138$-$097  &    1Jy                    &    LBL    &    0.733   &   -           &   -         &   -           &      no host&           &   \\
0219+428    &    1ES, 3C 66A             &    LBL    &    0.444   &   14.80
(0.05) &      26.2   &         0.10  &      36.3   &       0.59&      15  \\
0301$-$243  \footnote{Object observed by our team at the CASLEO
telescope equipped with the Turin photopolarimeter (Scaltriti et
al. 1999, Tommasi 2000, Tommasi et al. 2001, Treves et al.
1999).}&    PKS   &    LBL    &   0.260&   - &      8.2
&         0.54  &      7.7    &           &          \\
0414+009    &    HEAO-A3                &    HBL    &    0.287   &   17.05 (0.06)&      3.6    &         -     &      3.7    &           &       \\
0735+178    &    PKS                    &    LBL    &    $>$0.424&15.54 (0.05)
&     9.8     &         0.25  &      $>$35  &     0.41  &    17       \\
0823$-$223  &    PKS                    &    LBL    &    $>$0.910&   -
&      15.3   &         0.25  &      no host&           &          \\
0851+202    &    1Jy, OJ~287            &    LBL    &    0.306   &   16.72 (0.09)&      12.4   &         -0.28 &      $>$26  &           &          \\
2005$-$489 $\,^f$ &    PKS   &    HBL    &   0.071    &   -           &      4.3    &        0.31   &      5.2    &     1     &  15     \\
2155$-$304 $\,^f$ &    PKS   &    HBL    &    0.116   &   -           &      4.7    &         0.42  &      7.2    &    1.22   &   18     \\
2200+420    &    1Jy, BL Lac            &    LBL    &    0.069   &   13.69 (0.04)&      4.0    &     -0.27     &   6.1       &   0.43    &    11 \\
\hline
\end{tabular}
\end{minipage}
\end{table*}

The Turku polarimeter (Turpol) was used for the NOT observations.
This is a double channel chopping photopolarimeter, designed by
Piirola (\cite{Piirola}). Although the instrument conception is
rather old, it remains particularly suited to multifrequency
observations, thanks to its capability of performing simultaneous
photopolarimetry in five different bands. Polarization of the
light collected by the telescope is determined by a half-wave or
quarter-wave retarder plate (to measure linear or circular
polarization, respectively, with the highest efficiency), rotated
through 8 positions by 22.5$^\circ$ steps. Then a calcite slab
splits the light into ordinary and extraordinary rays that pass
through identical diaphragms. Both components of the diffuse
background from adjacent sky patches enter both diaphragms,
resulting in a cancellation of the polarization of sky light.
Finally, the beam is split by four dichroic+bandpass filter
combinations and sent to five photomultipliers. Each of these
channels reproduces the spectral response of one of the
\mbox{$UBVRI$} Johnson-Cousins bands. In this way, truly
simultaneous multiband observations can be performed. For all our
observations, an integration time of 10 seconds was used for each
position of the retarder plate, giving a polarimetric measurement
every 3.5 minutes. In order to improve the {\em signal to noise}
ratio, intranight measurements were then binned with four plate
cycles (about 15 minutes). A 10 second sky background integration
was normally performed every 15 minutes. High and null
polarization standard stars from Schmidt et al.
(\cite{Schmidt92}), observed several times per night, were used
to determine the instrumental parameters: instrument induced
polarization and orientation of the zero point of the retarder
plate with respect to the North.

Data reduction was performed using dedicated routines. They allow
calculation of polarization percentage ($P$) and position angle
($PA$) by fitting the counts recorded in the eight positions of
the retarder plate with a suitable cosine function using a
least-squares algorithm. Subtraction of an average sky value
obtained by interpolating between the nearest sky background
acquisitions taken close to the object is automatically performed
and the instrumental Stokes parameters are subtracted to obtain
the true polarization state from the measured one. Error estimates
take into account both the uncertainty from photon statistics and
from the least-squares fit. In particular, the larger of these two
contributions is adopted. In this way the error cannot be
underestimated, as it could be if only the fit uncertainty was
used. Binned points and nightly means can be calculated too, by
averaging Stokes parameters over single polarimetric measurements.

A summary of the results is reported in Table~\ref{poltab}
with 1$\sigma$ errors in brackets. The uncertainties for 3C 66A
and PKS~0735+178 during December 11 and 12 are the maximum between
statistical errors associated with the individual points and
standard deviations of intranight binned measurements. So,
intrinsic variability of the objects during those nights is taken
into account as a source of uncertainty on the nightly means.
\begin{table*}[h]
 \label{poltab} \centering
\begin{minipage}{388pt} \caption[]{Journal of Observations}
\centering \footnotesize
\begin{tabular}{ccccccc} \hline
\emph{Date (Dec 1999)} & \emph{JD} & \emph{Duration (min.)} &
\emph{Pol. state} $\,^a$ & \emph{Filter} & \emph{$P$ (\%)} &
\emph{$PA$ ($^{\circ}$)}  \\ \hline

\multicolumn{7}{c}{\bf 0138-097} \\*
  &&   &   & &             &            \\*
13&2451526&60& c &$U$& 0.57 (0.67)&  -\\*
  &&   &   &$B$& 0.32 (1.03)&  -\\*
  &&   &   &$V$& -0.14 (0.73)&  -\\*
  &&   &   &$R$& -0.20 (0.69)&  -\\*
  &&   &   &$I$& 1.41 (1.12)&  -\\[12pt]

\multicolumn{7}{c}{\bf 3C 66A (0219+428)} \\*
  &&   &   & &             &            \\*
 10&2451523&45& l &$U$& 28.53 (0.38)& 178.9 (0.4)\\*
  &&   &   &$B$& 27.32 (0.29)& 179.5 (1.2)\\*
  &&   &   &$V$& 27.27 (1.32)& 179.2 (1.2)\\*
  &&   &   &$R$& 26.52 (0.28)& 178.8 (1.5)\\*
\smallskip
  &&   &   &$I$& 25.09 (0.35)& 0.4 (2.1)\\

11&2451524&135 & l &$U$& 27.31 (0.22)& 1.7 (1.2)\\*
  &&   &   &$B$& 26.81 (0.33)&  1.2 (1.1)\\*
  &&   &   &$V$& 26.36 (0.35)&  1.1 (1.5)\\*
  &&   &   &$R$& 25.67 (0.91)&  1.0 (1.6)\\*
\smallskip
  &&   &   &$I$& 25.33 (0.53)&  1.4 (1.4)\\

12&2451525&75& l &$U$& 26.09 (0.69)& 2.1 (0.5)\\*
  &&   &   &$B$& 25.52 (0.79)& 2.9 (1.0)\\*
  &&   &   &$V$& 25.60 (0.45)& 2.9 (0.8)\\*
  &&   &   &$R$& 25.07 (0.47)& 1.5 (0.6)\\*
\smallskip
  &&   &   &$I$& 24.73 (0.84)& 3.1 (0.8)\\

13&2451526&90& c &$U$& 0.13 (0.10)& - \\*
  &&   &   &$B$& 0.09 (0.11)& - \\*
  &&   &   &$V$& 0.28 (0.14)& - \\*
  &&   &   &$R$& 0.20 (0.08)& - \\*
\smallskip
  &&   &   &$I$& 0.12 (0.14)& - \\

14&2451527&30& l &$U$& 27.95 (0.37)&  2.4 (0.6)\\*
  &&   &   &$B$& 26.66 (0.96)&  3.6 (1.6)\\*
  &&   &   &$V$& 26.80 (1.30)&  2.9 (1.0)\\*
  &&   &   &$R$& 25.45 (0.30)&  3.1 (0.8)\\*
  &&   &   &$I$& 25.00 (0.39)&  2.6 (1.2)\\[12pt]
\end{tabular}
\end{minipage}
\end{table*}
\addtocounter{table}{-1}
\begin{table*}[h!]
 \centering
\begin{minipage}{388pt}\caption[]{Continued}
\centering \footnotesize
\begin{tabular}{ccccccc} \hline
\emph{Date (Dec 1999)} & \emph{JD} & \emph{Duration (min.)} &
\emph{Pol. state} $\,^a$
& \emph{Filter} &
\emph{$P$ (\%)} & \emph{$PA$ ($^{\circ}$)}  \\ \hline

\multicolumn{7}{c}{\bf 0414+009} \\*
  &&   &   & &             &            \\*
14&2451527&75& l &$U$& 3.38 (0.51)&  139.4 (4.3)\\*
  &&   &   &$B$& 3.87 (0.69)&  151.6 (5.0)\\*
  &&   &   &$V$& 3.63 (0.89)&  150.8 (6.9)\\*
  &&   &   &$R$& 2.89 (0.64)&  133.1 (6.2)\\*
  &&   &   &$I$& 4.0 (1.2)&  147.1 (8.7)\\*[12pt]

\multicolumn{7}{c}{\bf 0735+178} \\*
  &&   &   & &             &            \\*
 10&2451523&15& l &$U$& 11.66 (0.41)& 135.7 (1.0)\\*
  &&   &   &$B$& 9.71 (0.45)& 133.2 (1.3)\\*
  &&   &   &$V$& 9.89 (0.61)& 136.5 (1.8)\\*
  &&   &   &$R$& 9.30 (0.38)& 132.5 (1.2)\\*
\smallskip
  &&   &   &$I$& 8.53 (0.54)& 138.7 (1.7)\\

11&2451524&135 & l &$U$& 12.00 (0.58)& 138.5 (2.3)\\*
  &&   &   &$B$& 10.72 (0.64)&  136.9 (3.3)\\*
  &&   &   &$V$& 10.28 (0.53)&  137.6 (3.5)\\*
  &&   &   &$R$& 9.66 (0.36)&  137.3 (3.0)\\*
\smallskip
  &&   &   &$I$& 9.52 (0.47)&  138.9 (4.8)\\

12&2451525&150& l &$U$& 10.56 (0.56)& 134.4 (2.8)\\*
  &&   &   &$B$& 9.36 (0.57)& 134.2 (3.0)\\*
  &&   &   &$V$& 9.36 (0.64)& 133.5 (3.4)\\*
  &&   &   &$R$& 8.85 (0.43)& 133.7 (1.6)\\*
\smallskip
  &&   &   &$I$& 8.95 (1.04)& 134.4 (3.7)\\

13&2451526&105& c &$U$& 0.04 (0.14)& -\\*
  &&   &   &$B$& 0.05 (0.12)& -\\*
  &&   &   &$V$& 0.02 (0.15)& -\\*
  &&   &   &$R$& 0.01 (0.07)& -\\*
\smallskip
  &&   &   &$I$& 0.04 (0.13)& - \\

14&2451527&30& l &$U$& 12.12 (0.28)&  122.6 (0.7)\\*
  &&   &   &$B$& 11.84 (0.62)&  123.7 (0.8)\\*
  &&   &   &$V$& 11.62 (0.78)&  122.0 (1.1)\\*
  &&   &   &$R$& 10.33 (0.83)&  125.6 (0.8)\\*
  &&   &   &$I$& 9.90 (0.71)&  124.8 (2.9)\\[12pt]
\end{tabular}
\end{minipage}
\end{table*}
\addtocounter{table}{-1}
\begin{table*}[h!]
 \centering
\begin{minipage}{388pt}\caption[]{Continued}
\centering \footnotesize
\begin{tabular}{ccccccc} \hline
\emph{Date (Dec 1999)} & \emph{JD} & \emph{Duration (min.)} &
\emph{Pol. state} \footnote{Polarization state: l=linear
polarization; c=circular polarization.} & \emph{Filter} &
\emph{$P$ (\%)} & \emph{$PA$ ($^{\circ}$)}  \\ \hline

\multicolumn{7}{c}{\bf 0823-223} \\*
  &&   &   & &             &            \\*
 10&2451523&45& l &$U$& 16.81 (0.84)& 7.2 (1.4)\\*
  &&   &   &$B$& 14.78 (0.59)& 9.1 (1.2)\\*
  &&   &   &$V$& 15.64 (0.89)& 4.1 (1.6)\\*
  &&   &   &$R$& 14.42 (0.47)& 4.1 (0.9)\\*
\smallskip
  &&   &   &$I$& 12.43 (0.76)& 2.2 (1.8)\\

13&2451526&82& c &$U$& 0.11 (0.18)& -\\*
  &&   &   &$B$& 0.43 (0.20)& -\\*
  &&   &   &$V$& 0.18 (0.18)& -\\*
  &&   &   &$R$& 0.01 (0.14)& -\\*
\smallskip
  &&   &   &$I$& 0.26 (0.21)& - \\

14&2451527&45& l &$U$& 15.76 (0.40)&  12.7 (0.7)\\*
  &&   &   &$B$& 15.51 (0.34)&  12.3 (0.6)\\*
  &&   &   &$V$& 15.01 (0.48)&  11.9 (0.9)\\*
  &&   &   &$R$& 13.51 (0.28)&  10.3 (0.6)\\*
  &&   &   &$I$& 12.88 (0.41)&  7.8 (0.9)\\[12pt]

\multicolumn{7}{c}{\bf OJ287 (0851+203)} \\*
  &&   &   & &             &            \\*
14&2451527&37& l &$U$& 11.27 (0.72)&  161.4 (1.8)\\*
  &&   &   &$B$& 11.45 (0.65)&  163.5 (1.6)\\*
  &&   &   &$V$& 12.40 (0.85)&  162.4 (2.0)\\*
  &&   &   &$R$& 14.23 (0.52)&  162.0 (1.0)\\*
  &&   &   &$I$& 13.23 (0.80)&  161.6 (1.7)\\[12pt]

\multicolumn{7}{c}{\bf $B$L Lac (2200+420)} \\*
  &&   &   & &             &            \\*
14&2451527&30& l &$U$& 3.66 (0.30)&  19.8 (2.4)\\*
  &&   &   &$B$& 4.13 (0.19)&  15.2 (1.3)\\*
  &&   &   &$V$& 4.03 (0.20)&  14.2 (1.4)\\*
  &&   &   &$R$& 4.29 (0.09)&  11.1 (0.6)\\*
  &&   &   &$I$& 4.80 (0.10)&  8.4 (0.6)\\[12pt]
\hline
\end{tabular}
\end{minipage}
\end{table*}

Absolute flux measurements from the counts recorded by the
photopolarimeter are much more sensitive than Stokes parameter
measurements (involving differences between close count
recordings) to sky background/transparency fluctuations and seeing
effects, as well as to small pointing or guiding errors,
especially if relatively low-brightness targets are observed. As a
consequence, no useful simultaneous photometry could be extracted
from our polarimetric data. Therefore, we used the best $V$-band
images taken with the standby CCD camera (Stancam) during the
pointing phase to each source to obtain information on the
photometric state of the BL Lacs during our observations. Results
are reported in Table~\ref{sample}.
The magnitude
differences measured between the standard stars included in the
field of view are very similar to the standard $\Delta V$ values
reported in the literature (Smith \& Sitko \cite{Smith91};
Fiorucci \& Tosti \cite{Fiorucci96}; Fiorucci et al.
\cite{Fiorucci98}). Differential photometry yielded a precision
of about $\pm$0.02~mag, but the $V$ magnitudes of the comparison
stars are known with a greater error (about $\pm$0.06~mag).
Hence, the uncertainties reported in brackets are mainly due to
the reduction to the standard photometric system.

\section{Results}
\subsection{Linear Polarization}
\subsubsection{3C 66A}
\begin{figure*}
\resizebox{12cm}{!}{\includegraphics{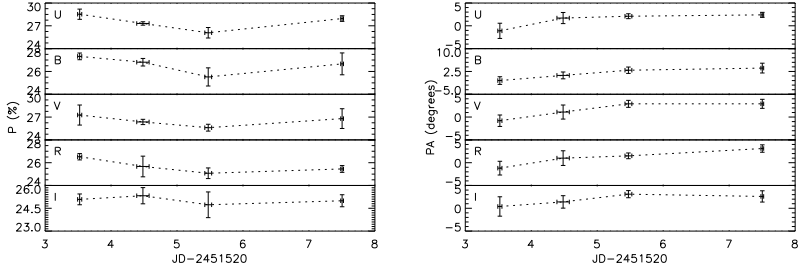}} \hfill
\parbox[b]{55mm}{ \caption{Nightly mean values of $P$ (left panels) and
$PA$ (right panels) in $UBVRI$ bands for 3C 66A. Horizontal bars
represent the duration of the observations.} \label{0219inter}}
\end{figure*}

\begin{figure*}
\resizebox{12cm}{!}{\includegraphics{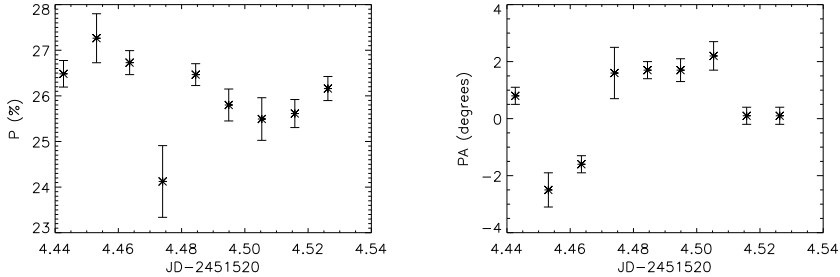}} \hfill
\parbox[b]{55mm}{\caption{Intranight behaviour of 3C 66A during December 11, 1999
in $R$ band with 15 minute time bins.} \label{0219intran}}
\end{figure*}

Our results on this source, showing an average polarization of
27.3\% in $U$ band, confirm that it is one of the most strongly
polarized BL Lacs (Takalo \& Sillanp\"a\"a \cite{Takalo93}). Some
day-to-day variability of both $P$ and $PA$ is present
(Fig.~\ref{0219inter}).
A simple $\chi^2$ test shows significant variability of $P$ at a
95\% confidence level or more in $U$, $B$ and $R$ filters and of
$PA$ in $U$, $V$ and $R$. Moreover, variations in the five bands
are well correlated. In Fig.~\ref{0219intran} polarization and
position angle during the night of December 11 in the $R$ filter
are reported, with 15 minute binning.
Variations are detected both in polarization and position angle,
with higher significance for $PA$. The sharp minimum in $PA$ is
noticeably lasting for about half an hour in the first part of the
night and recorded also in the other bands, while variations in
$P$ are not significant in $U$, $B$, $V$ and $I$ filters.

\begin{figure*}
\resizebox{12cm}{!}{\includegraphics{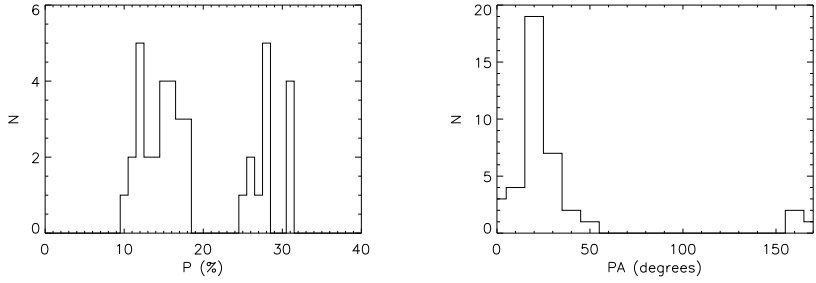}} \hfill
\parbox[b]{55mm}{\caption{Histograms of literature data for the $V$ band
polarimetry of 3C 66A (our results are included).}
\label{lett0219}}
\end{figure*}

A comparison of our results with the polarization measurements
reported in the literature
(Mead et al. \cite{Mead}; Sitko et al. \cite{Sitko85}; Smith et
al. \cite{Smith87}; Takalo \cite{Takalo91}; Takalo \&
Sillanp\"a\"a \cite{Takalo93}; Takalo et al. \cite{Takalo94};
Valtaoja et al. \cite{Valtaoja91}) is given in
Fig.~\ref{lett0219}.
All the collected data are reported with the same weight, even if
observations do not uniformly cover the spanned time period; most
of them have been performed in short campaigns separated by long
intervals during which the source was not observed. The $P$
values are rather scattered, and seem to concentrate in
two broad regions of the plot: one between 10\% and 18\%, the
second between 25\% and 31\%. However, the number of available
measurements is too low to claim a real bimodal distribution.
$PA$ shows a strongly preferred value between 20$^\circ$ and
30$^\circ$. No correlation between $P$ and $PA$ was found.

\subsubsection{PKS~0735+178}
\begin{figure*}
\resizebox{12cm}{!}{\includegraphics{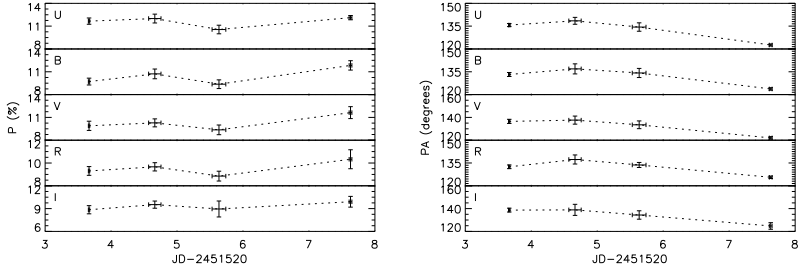}} \hfill
\parbox[b]{55mm}{\caption{Nightly mean values of $P$ (left panels) and $PA$ (right
panels) in $UBVRI$ bands for PKS~0735+178. Horizontal bars
represent the duration of the observations.} \label{0735inter}}
\end{figure*}

\begin{figure*}
\resizebox{12cm}{!}{\includegraphics{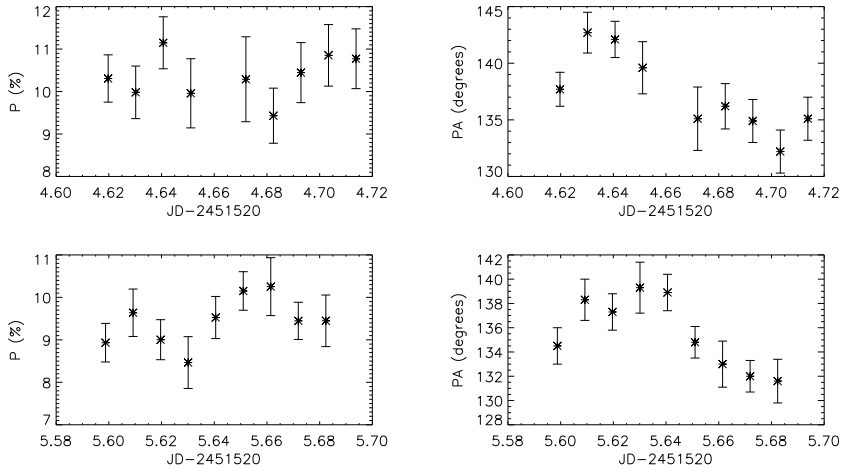}} \hfill
\parbox[b]{55mm}{\caption{Intranight behaviour of PKS~0735+178 during December 11
in $V$ band (upper panels) and December 12 in $B$ band with 15
minute time bins.} \label{0735intran}}
\end{figure*}

Nightly averages of $P$ and $PA$ measurements for this source are
reported in Fig.~\ref{0735inter}.
Little variability is recorded in polarization percentage
(significance level greater than 95\% only in the $V$ filter),
and more significant variations in $PA$ are seen. In all bands, an
increase in $P$ is seen in the last night, while the $PA$
simultaneously rotates significantly with respect to the previous
nights. Also on intranight time scales, more significant
variations are seen in position angle than in polarization
percentage. For the two nights with longest monitoring, we report
in Fig.~\ref{0735intran} the $P$ and $PA$ curves in the filters
where the variations are most apparent. Similar trends are seen
also in the other bands.

\begin{figure*}
\resizebox{12cm}{!}{\includegraphics{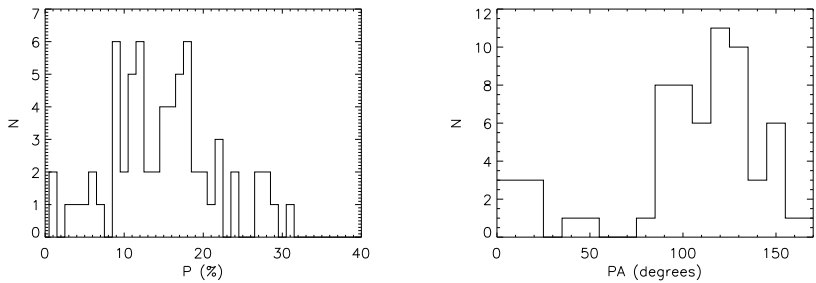}} \hfill
\parbox[b]{55mm}{\caption{Histograms of literature data for the $V$ band
polarimetry of PKS~0735+178 (our results are included).}
\label{lett0735}}
\end{figure*}

Also for this BL Lac object, we searched the literature for
previous polarimetric data (Brindle et al. \cite{Brindle}; Mead
et al. \cite{Mead}; Puschell et al. \cite{Puschell83}; Sitko et
al. \cite{Sitko85}; Smith et al. \cite{Smith87}; Takalo
\cite{Takalo91}; Takalo et al. \cite{Takalo92}; Valtaoja et al.
\cite{Valtaoja91}, \cite{Valtaoja93}) and reported them in
histograms along with our data (Fig.~\ref{lett0735}) as done in
the case of 3C 66A.
Data are more sparsely distributed with respect to those of that
object, but also in this case a broad peak in $PA$ between
90$^\circ$ and 160$^\circ$ is suggested. The source showed very
different levels of $P$ in the past years, covering the whole
range from about 1\% to more than 30\%.

\subsubsection{Other BL Lac objects} Linear polarization measurements also were performed
for four other BL Lac objects: H~0414+009, PKS~0823$-$223, OJ287
and BL Lac (see Table~\ref{poltab}).
PKS~0823$-$223 was observed twice and found to show different
levels of polarization in December 10 and 14. While $P$ was
basically constant, $PA$ rotated by about 3$^\circ$--8$^\circ$ in
the different filters. The other listed sources were observed
sparsely during the run for short intervals, so no variability
analysis could be performed.

Comparison with literature data shows that the three sources
H~0414+009 (Impey \& Tapia \cite{Impey88}; Mead et al.
\cite{Mead}), OJ~287 (e.g. Smith et al. \cite{Smith87};
Sillanp\"a\"a et al. \cite{Sillanpaa91}; Takalo et al.
\cite{Takalo92}) and BL Lac (e.g. Puschell et al.
\cite{Puschell83}; Moore et al. \cite{Moore87}; Sillanp\"a\"a et
al. \cite{Sillanpaa93}) were found at polarization levels comparable to
their historical values. In particular, the last two
sources showed large variations in the past both in $P$ and $PA$
and our values lie well within the observed variability ranges.
PKS~0823$-$223 was found at a different polarization level with
respect to that recorded at previous epochs. It has been observed
in the past with $P$ ranging from 4.0\% to 12.7\% at wavelengths
around 5000~\AA~ (Impey \& Tapia \cite{Impey88}; Wills et al.
\cite{Wills92}; Visvanathan \& Wills \cite{Visvanathan98}), while
during our campaign it was polarized up to 15.6\% in $V$ band.
However, only few past measurements are available, so the level of
polarization detected by us may be not that rare.

\subsection{Wavelength dependence of polarization}
Wavelength dependence of polarization (WDP) was investigated.
Following Smith \& Sitko (\cite{Smith91}), we introduced the
parameters
\begin{equation}
P_\nu=\frac{d\log P}{d\log \nu} \label{pnu}
\end{equation}
and
\begin{equation}
{\it PA}_\nu=\frac{d {\it PA}}{d\log \nu} \, ,
\end{equation}
which describe the spectral shape of $P$ and $PA$. They were
determined by fitting power-laws to the nightly means of the
$UBVRI$ linear polarization percentages and position angles,
considered as monochromatic measurements at the effective
wavelength of the filters. In Table~\ref{tabWDP}
we report the most statistically significant values of $P_\nu$
and ${\it PA}_\nu$ obtained for some objects/nights, with their
standard deviations resulting from the fit.
\begin{table}  \label{tabWDP}
\centering
\begin{minipage}{250pt}\caption[]{Results from WDP analysis.}
\centering \footnotesize
\begin{tabular}{ccccc}
 \hline {\em Object} &{\em Date} & {\em JD}
& {$P_\nu$} & {${\it PA}_\nu$} \\
 & {\em (Dec 99)} & & & \\\hline

3C 66A& 10&2451523&0.11$\pm$0.02& - \\*
     & 11&2451524&0.08$\pm$0.02& - \\*
     & 12&2451525&0.06$\pm$0.04& - \\*
     & 14&2451527&0.13$\pm$0.02& - \\*[12pt]
0735+178& 10&2451523&0.29$\pm$0.06& - \\*
            & 11&2451524&0.26$\pm$0.07& - \\*
            & 12&2451525&0.21$\pm$0.09& - \\*
            & 14&2451527&0.21$\pm$0.06& -3.6$\pm$1.5 \\*[12pt]
0823+223    & 10&2451523&0.28$\pm$0.08& 5.6$\pm$2.4 \\*
            & 14&2451527&0.24$\pm$0.04& 4.9$\pm$1.0 \\*[12pt]
OJ287       & 14&2451527&-0.28$\pm$0.08& - \\* [12pt]

BL Lac & 14&2451527&-0.27$\pm$0.05& 10.7$\pm$1.6 \\* \hline
\end{tabular}
\end{minipage}
\end{table}
In most cases ${\it PA}_\nu$ is poorly determined and has been
omitted in the table. H~0414+009 does not show significant
wavelength dependence of polarization. While most objects show
the usual trend with $P$ decreasing with decreasing frequency
($P_\nu
> 0$), that can be partially due to dilution by the host galaxy
light, noticeably two sources, namely BL Lac and OJ~287, show
$P_\nu < 0$. Similar results have been obtained for these two
objects in past observations, see e.g. Sillanp\"a\"a et al.
(\cite{Sillanpaa91}, \cite{Sillanpaa93}). For the sources
observed more than once during the campaign, no significant
variability of the WDP can be evidenced.

\subsection{Circular polarization}
Circular polarization was investigated during the night of
December 13, 1999 on four BL Lac objects (1Jy~0138$-$097, 3C 66A,
PKS~0735+178 and PKS~0823$-$223) and the results are reported in
Table~\ref{poltab}.
Only in 3C 66A, the object with the smallest errors due to its
relatively high brightness, was circular polarization marginally
detected at the 2$\sigma$ level in $V$ and $R$ bands. A detection
of non-null circular polarization in 3C 66A was claimed by Takalo
\& Sillanp\"a\"a (\cite{Takalo93}), with larger values than found
in our data. We conservatively put an upper limit of 0.4\% with a
3$\sigma$ confidence level on the circular polarization of this
BL Lac. In the case of PKS~0735+178, marginal detection of
circular polarization $P=-0.24 \rm{\%}\pm 0.12$\% was reported by
Valtaoja et al. (\cite{Valtaoja93}). However, no such
polarization was recorded during our campaign, even if our
uncertainty level is similar. Our data show an upper limit of
about 0.4\% at the 3$\sigma$ level for PKS~0735+178. No
significant circular polarization was found for the two other
observed BL Lacs. Upper limits can be fixed at 2.5\% and 0.5\%
for 1Jy~0138$-$097 and PKS~0823$-$223, respectively.

\section{Discussion}
The two sources which have been intensively monitored within this
programme, 3C 66A and PKS~0735+178, exhibit a modest internight
variability, which is more apparent in the position angle. This
occurs also on an intranight scale and the overall behaviour is
similar to that observed in B2~1308+326 (Puschell et al.
\cite{Puschell79}), BL Lac (Moore et al. \cite{Moore87}),
S5~0716+174 (Impey et al. \cite{Impey00}) and PKS~2155$-$304
(Tommasi et al. \cite{Tommasi01}). In the last three cases, total
integration times were larger by an order of magnitude and more
extreme variability episodes were recorded. As discussed by
Tommasi et al. (\cite{Tommasi01}), a possible interpretation is
that the emitting region is a relativistic jet, containing
substructures of different polarization, with variable
intensities.

Our results confirm the finding of a preferred polarization level
over a time scale of years, indicating an underlying stability of
the jet structure, even if in the presence of variability of the
single subcomponents. In particular, the existence of a preferred
position angle (see e.g. Jannuzi, Smith \& Elston
\cite{Jannuzi94}) could be related to a regular large-scale
component of the magnetic field (Tommasi et al. \cite{Tommasi01}).

A summary of the results of the present observations is reported
in Table~\ref{sample},
together with data on some sources that have been observed
previously by our team. We have also reported the core dominance
($CD$) parameter in the $R$ band, i.e. the ratio between the
nuclear intensity to that of the host galaxy, collected from the
literature (Wurtz et al. \cite{Wurtz96}; Scarpa et al.
\cite{Scarpa00}). It is apparent that the WDP is different for
the various sources. Although the $P_\nu$ parameters should be
corrected for dilution due to the host galaxy light (which in a
first approximation can be considered as unpolarized), this
difference holds also when comparing only the most core-dominated
objects, for which the galaxy correction should be negligible.
This may be ascribed to the different spectral appearance of LBL
and HBL: with one exception (PKS 0301--243), the LBL in our
sample have the lower values of $P_\nu$ (i.e., steeper polarized
spectra) while the WDP for the HBL is flatter (although not
significantly saw within the limited number statistics). This
reflects the fact that, when the galaxy correction is negligible,
the shape of the optical polarization should be that of ``pure"
synchrotron emission and therefore should depend on the position
of the synchrotron peak (i.e., more HBL-like sources should have a
flatter slope).

In two LBL in our sample, OJ 287 and BL Lac, the $P_\nu$ slope is
negative, meaning that the polarization increases with
wavelength. This is the opposite of what is expected for pure
synchrotron radiation. Since the effect is significant, we argue
that a source of thermal nature, like an accretion disk, may
dilute the non-thermal emission of these objects and determine
the observed wavelength dependence (see Wandel \& Urry
\cite{wanurr91}; Smith \& Sitko \cite{Smith91}). In BL Lac, an
accretion disk has indeed been proposed by Corbett et al.
(\cite{Corbett96}) and Corbett et al. (\cite{Corbett00}) as the
powering source of the observed, variable H$\alpha$ emission
line.  The presence of an accretion disk in OJ~287 has been
invoked to explain the observed flux and spectral variability
(Sillanp\"a\"a et al. \cite{sill88}; Kidger et al.
\cite{kidger91}; Katz \cite{katz97}).  However, the $P_\nu$ of OJ
287 in past polarimetric monitorings in optical bands has not
been systematically negative (see e.g. Takalo et al.
\cite{Takalo92}), and, overall, the source exhibited
preferentially a WDP behavior opposite to the one observed by us,
namely decreasing polarization with increasing wavelength (Takalo
\cite{takalo94}). One may argue that, since the accretion disk
component is presumably more modestly variable than the beamed
non-thermal one, it emerges during phases of quiescent
synchrotron emission, so that a negative $P_\nu$ value may be
more often seen during low brightness states of the source, such
as the one we have monitored in our campaign.

\begin{figure}
\resizebox{\hsize}{!}{\includegraphics{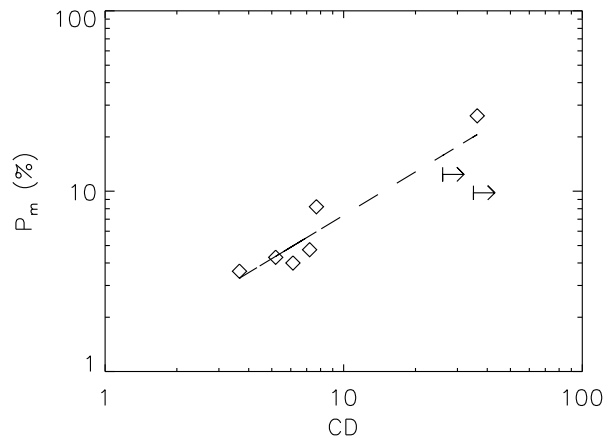}}
\caption{Polarization vs. core dominance ($CD$) for the objects
in the sample. The dashed line represents the fit over the points
with $CD\leq 10$, extrapolated towards higher values of $CD$.
The arrows represent lower limits on the $CD$ for
PKS~0735+178 and OJ~287.} \label{CD}
\end{figure}

Following Wurtz et al. (\cite{Wurtz96}) we have plotted the
average optical polarization in the $V$ band $P_m$ vs. $CD$ in
Fig.~\ref{CD}.
At the lower values, say $CD \leq 10$, there may be a monotonic
growth, which could result only in part from the effect of
dilution by the host galaxy. It has been highlighted in the
figure by tracing the line representing a linear regression over
those points in logarithmic scales. The fact that 3C 66A (the
object with the highest polarization recorded) lies on the line
is noticeable and confirms that dilution is not the main cause of
the observed dependence, since in this highly polarized and
strongly core-dominated source, the host galaxy light should play
little part. Even if the evidence of a $P$ vs. $CD$ correlation is
based on few points, it interestingly confirms the suggestion of
Wurtz et al. (\cite{Wurtz96}), who found a correlation between
two quantities at a 99\% probability level. The result is not
easy to interpret, and could suggest a correlation between
polarization and beaming properties, which may manifest itself in
the observed CD.  A dependence of the observed polarization
percentage from the degree of jet collimation and radiation
beaming is expected, based on recent speculation made in the
context of polarized afterglow emission from Gamma-Ray Bursts
(Gruzinov \cite{Gruzinov99}; Gruzinov \& Waxman \cite{gruzwax99};
Covino et al. \cite{Covino99}; Wijers et al. \cite{Wijers99}). On
the other hand, we note that the CD parameter as defined here
could reasonably be associated with the radio nuclear-to-extended
emission, which is a clear tracer of beaming in BL Lacs (see Fig.
5 in Ghisellini et al. \cite{gg93}).  Regrettably, we have
insufficient statistics to prove the correlation of optical and
radio CD in our sample (only 4 of our objects have a measured
radio CD), and the issue is in any case sensitive to nuclear
variability, which is stronger in optical than in radio. We have
therefore considered a model of the overall spectrum of blazars
from radio to gamma ray frequencies, where a plasma radiates
mainly via synchrotron and inverse Compton mechanisms. The
relativistic motion of the plasma toward the observer is
characterized by the beaming factor $\delta$. Our sample contains
only BL Lac objects, and therefore the scattering of the
relativistic electrons should occur mainly on the synchrotron
photons, and the spectrum should be represented by the so-called
synchrotron self-Compton model (SSC). The observation of the
overall spectrum and of its time variability allows us to
constrain the various parameters which enter the model (e.g.
Ghisellini et al. \cite{Ghisellini98}). In particular one can
determine $\delta$ and the magnetic field intensity B, which a
priori should correlate directly with polarization properties.
Their values are reported in Table~\ref{sample}
(Ghisellini et al. \cite{Ghisellini98}; Tagliaferri et al.
\cite{Tagliaferri01}). Besides the well-known fact that LBL are
more polarized than HBL (see e.g. Jannuzi et al.
\cite{Jannuzi94}; Ulrich et al. \cite{Ulrich97}), we notice that
HBL appear to be associated with a stronger magnetic field in our
sample. However, no clear direct dependence of $P$ on $\delta$ is
apparent. Obviously, the objects under consideration are rather
few. A search for a possible correlation of the circular
polarization with $B$ or $\delta$ would be hazardous, because of
the large uncertainties on $P$.

The investigation of polarization of BL Lac objects can give a
large amount of information that could help to constrain jet
models (Jones \cite{jones88}; Bj\"ornsson \cite{bjorn93}).
Coordinated polarimetric monitoring at optical, near-infrared and
radio wavelengths has allowed several authors to elucidated the
structure and distribution of the magnetic field, the interplay
of different emission components and the location of emitting
regions in blazars thanks to the variability of multiwavelength
polarization percentage (e.g., Brindle et al. \cite{Brindle};
Kikuchi et al. \cite{kikuchi88}; Cawthorne \& Wardle
\cite{cawthorne88}; Sillanp\"a\"a \cite{sill91}; Gabuzda \&
G\'omez \cite{Gabuzda01}). Therefore other intensive campaigns
seem strongly needed. In particular, simultaneous observations of
the optical polarization and X-ray flux could better investigate
the physical environment of the inner part of the synchrotron
source. Moreover, simultaneous polarimetric and photometric data
in the optical, best obtained with CCD-based instruments, could
help to clarify how the injection of relativistic particles
correlates with variations in the magnetic field.

\begin{acknowledgements} We thank the NOT staff for technical support
during the observations. Financial support from EC grant ERBFMRXCT
98-0195 and Italian MURST COFIN 98021541 are acknowledged.
\end{acknowledgements}

\clearpage
\end{document}